\documentclass[aps,pre,twocolumn,showpacs,superscriptaddress]{revtex4}
\usepackage{graphicx}
\usepackage{dcolumn}
\usepackage{bm}
\usepackage[hang, nooneline]{subfigure}
\usepackage[usenames,dvipsnames]{color} 

\usepackage{setspace}
\usepackage[latin1]{inputenc}           
\usepackage[british]{babel}
\usepackage[T1]{fontenc}
\usepackage{color}
\usepackage{amssymb}
\usepackage{amsmath}
\usepackage[squaren]{SIunits}
\usepackage{multirow}                   
\usepackage{booktabs}

\usepackage{color}
\usepackage[normalem]{ulem}

\usepackage{graphics}
\usepackage[all]{xy}

\begin{document}


\title{
Predicting patterns for molecular self-organization on surfaces using interaction-site models  
} 

\author{Marta Balb\'as Gambra}
\affiliation{Arnold Sommerfeld Center for Theoretical Physics (ASC) and Center for
Nano Science (CeNS), Fakult{\"a}t f{\"u}r Physik,
Ludwig-Maximilians-Universit{\"a}t M{\"u}nchen, Theresienstra{\ss}e 37, 80333 M\"unchen, Germany}

\author{Carsten Rohr} 

\affiliation{Walther--Mei\ss ner Institute
for Low Temperature Research and Center for NanoScience, Department of Physics,
Ludwig-Maximilians-Universit{\"a}t M{\"u}nchen, Walther--Mei{\ss}ner Str. 8,  85748 Garching, Germany}

\author{Kathrin  Gruber}
\affiliation{Walther--Mei\ss ner Institute
for Low Temperature Research and Center for NanoScience, Department of Physics,
Ludwig-Maximilians-Universit{\"a}t M{\"u}nchen, Walther--Mei{\ss}ner Str. 8,  85748 Garching, Germany}

\author{Bianca Hermann}  
\affiliation{Walther--Mei\ss ner Institute
for Low Temperature Research and Center for NanoScience, Department of Physics,
Ludwig-Maximilians-Universit{\"a}t M{\"u}nchen, Walther--Mei{\ss}ner Str. 8,  85748 Garching, Germany}


\author{Thomas Franosch}
\affiliation{Institut f\"ur Theoretische Physik, Friedrich-Alexander-Universit\"at
Erlangen-N\"urnberg, Staudtstra{\ss}e 7, 91058 Erlangen, Germany}
\affiliation{Arnold Sommerfeld Center for Theoretical Physics (ASC) and Center for
Nano Science (CeNS), Fakult{\"a}t f{\"u}r Physik,
Ludwig-Maximilians-Universit{\"a}t M{\"u}nchen, Theresienstra{\ss}e 37, 80333 M\"unchen, Germany}

\date{\today}

\begin{abstract}Molecular building blocks interacting at
the nanoscale 
 organize spontaneously into stable
monolayers that display intriguing long-range ordering motifs on the
surface of  atomic substrates. The patterning process, if
appropriately controlled, represents a viable route to manufacture
practical nanodevices. With this goal in mind, we seek to capture
the salient features of the self-assembly process by means of an
\emph{interaction-site model}. The geometry of the building
blocks, the symmetry of the underlying substrate, and the strength
and range of interactions encode the self-assembly process. By
means of Monte Carlo simulations, we have predicted an ample
variety of ordering motifs which nicely reproduce the experimental
results. Here, we explore in detail  the phase behavior of the
system in terms of the temperature and the lattice constant of the
underlying substrate.
Our method is suitable to investigate the stability of the emergent patterns as
well as to identify the nature of the melting transition monitoring
appropriate order parameters.
\end{abstract}
%


\maketitle

\section{Introduction}

Molecular building blocks interacting at the nanoscale  organize
spontaneously into stable ordered  monolayers~\cite{Bartels:2010}. The patterns that
emerge upon  self-organization of simple supramolecular units
display a variety of symmetries and local ordering motifs with different  
degrees of packing  on the surface. Currently such surface coatings are
under active research due to their potential applications in
nanoscience, such as surface functionalization~\cite{barth2005engineering}, sensor surfaces~\cite{lehn2002toward}, or
molecular electronics~\cite{Joachim:2000}. 
The theoretical challenge is to provide tools that allow to predict
the patterns without performing the actual measurements for the
supramolecular units~\cite{Tomba:2009}. 
Most approaches rely on atomistic modeling where the stable
conformations are calculated using experimental measurements of the
molecular conformations on the surface as input. There, the molecular
subunit is resolved in all chemical details,   resulting in a large 
number of degrees of freedom that have to be included in the modeling.
 In molecular mechanics calculations (MM)~\cite{shi2006role} and molecular dynamics
 simulations (MD)~\cite{Chen:2008,Ilan:2008} the molecular interactions are parametrized by
 semi-empirical force fields which requires  a significant number of
 additional parameters. In MM the starting conformations are
 energy-minimized resulting in a stable configuration, yet the method  
intrinsically provides only the closest local minimum. In contrast, MD
simulations sample  different minima and incorporate, in principle, also
dynamical reconfigurational processes. The quantum aspects such as the
electronic density of states (DOS) measurable in scanning tunneling
microscopy (STM) can be computed using density functional theory 
(DFT)~\cite{Ilan:2008,glowatzki2008soft}, yet these calculations are limited to small systems.

Recently, a complementary approach has been introduced  employing coarse-grained
models which aim at predicting certain features of the patterns with 
the benefit of a great reduction  of complexity. For example, using an
effective hamiltonian accounting for the energetics of the respective
orientations of neighboring molecules, local ordering motifs for
oligopyridine supramolecules on a surface have been successfully
reproduced~\cite{breitruck2009short}. For mixtures of melamines, 
PTCDI, and PTCDA structural stability diagrams of two competing patterns have
been reported~\cite{silly2008deriving,weber2008role} using effective pair energies
between neighboring molecules on a lattice. For the case of flexible organic molecules, 
we have recently shown that a multi-method approach combining MM, DFT, and Monte Carlo simulations 
is capable of providing reliable predictions for the emerging multiple coexisting patterns on surfaces~\cite{rohr-molecular,Hermann:2010,Rohr:2010}.

In this paper we provide a detailed discussion  of the theoretical aspects of the use of interaction-site models for Fr\'echet dendrons. 
In particular, we discuss how salient features of the molecular structure can be identified and used to tailor a suitable coarse-grained model and then 
be studied by means of Monte-Carlo methods. Furthermore, we show that a phase diversity naturally emerges in agreement with experimental studies and discuss the corresponding phase stability as the packing fraction on the substrate or the temperature is varied. Then we provide a thorough discussion on the merits of the
approach and its limitations as well as possible extensions.

\section{Experimentally observed patterns}
\begin{figure}[h]
\centering
\resizebox{0.35\textwidth}{!}{%
\includegraphics{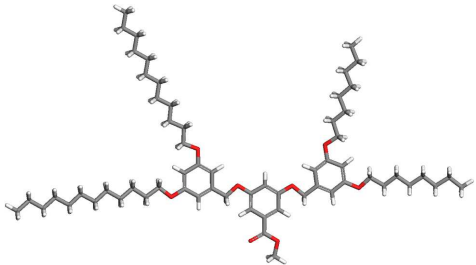}}
\caption{Stick model of the asymmetric Fr\'echet dendron used in the
experiments.} \label{fig:dendrons_chemical}
\end{figure}

In this Section we first summarize the experimental findings on molecular self-organization for Fr\'echet dendrons~\cite{Hawker:1990a,Hawker:1990b}.     
These second generation Fr\'echet dendrons are flexible supramolecules, which
consist of three phenyl rings symmetrically disposed at the
vertices of a triangle with two alkoxychains attached at each
lateral phenyl ring.
The arms of these chains consist  at one
side of the molecule of twelve  carbon
atoms, and on the other side  of 
eight carbon atoms---see the structural formula in Fig.~\ref{fig:dendrons_chemical}.

\begin{figure}[h]
\includegraphics[width=0.45\textwidth]{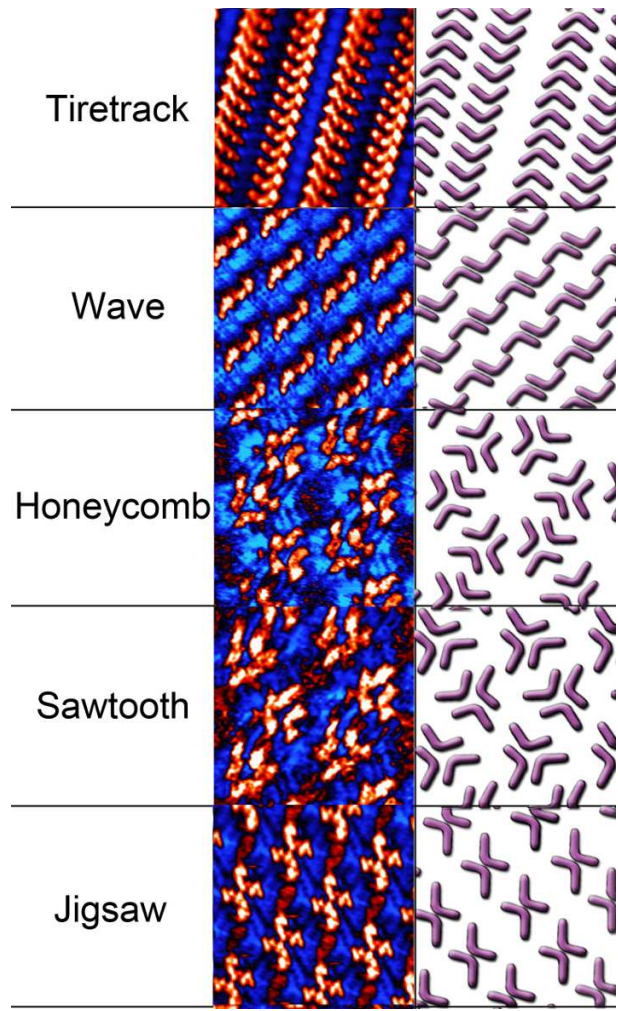}
\caption{Five experimentally observed patterns for Fr\'echet
  dendrons. STM images on the left and the corresponding schematic ordering motifs on the right.}
\label{fig:experiment_monolayers}
\end{figure}

Scanning tunneling microscopy (STM) images show that a large variety of 
self-organized ordering motifs on top of highly
oriented pyrolytic graphite (HOPG) emerges~\cite{Hermann:2006-selforganized,Constable:2007}.
This substrate exhibits a honeycomb-like
surface with a six-fold rotational symmetry.
Seven general ordering motifs have been reported, five  of which are flat-lying configurations \cite{rohr-molecular}. 
The tiretrack, wave, honeycomb, sawtooth, and jigsaw motifs are
displayed in Fig.~\ref{fig:experiment_monolayers}. These patterns coexist on the HOPG surface and phase
transform over time into the thermodynamically stable tiretrack
pattern. The schematics of the ordering motifs, Fig.~\ref{fig:experiment_monolayers}, have been obtained employing atomistic modeling
 by
molecular mechanics energy minimizations using the Forcite module of
Materials studio 4.3 and employing a universal force field. 

\section{The interaction-site model}

To obtain reliable predictions of the emergent pattern in the assembly
of monolayers without any
\emph{a priori} knowledge from the experiments, a drastic reduction of
complexity is highly desirable. Here, we  propose an
\emph{interaction-site model} which aims to provide a suitable
description of the self-organization process. The basic idea is to  reduce the numerous
microscopic forces to a few representative interactions acting on groups of
selected points. The major challenge consists of appropriately
identifying   the positions of the interaction centers and forces relevant for the self-organization.
We will demonstrate in this paper that the geometry of the building blocks,
the symmetry of the substrate, and the coupling of both are key
elements in the self-organization of monolayers. We will show for an
experimental model based on Fr\'{e}chet dendrons, that the
interaction-site model properly captures the essence  of the system,
predicts the observed patterns, as well as the temperature regime at
which they are stable. Moreover, monitoring suitable order parameters,
we are able to identify in principle the nature and location of the melting transition.   

\subsection{Construction of the interaction sites}

\begin{figure}[h]
\centering
\resizebox{0.45\textwidth}{!}{%
\includegraphics{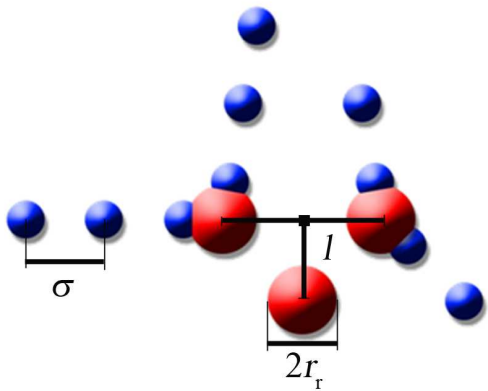}}
\caption{Interaction-site model for a  Fr\'echet dendron in conformation $\alpha$. Large spheres
  account for aromatic rings, whereas the
  small spheres represent
  subunits in the carbonated chains.
  Every sphere represents  four
  $\text{CH}_2$ units in the arms of the dendrons.}
\label{fig:interaction_site}
\end{figure}

The Fr{\'e}chet dendron introduced above is a suitable and especially
versatile model system to asses the
validity of an  interaction-site approach in the prediction of
self-organized monolayers.
According to the experimental observations the crucial features leading to
the self-organization are the steric repulsion between the molecular rings, the
weak interactions of the carbonated chains, as well as the coupling of the
building blocks with the substrate.
We model the symmetric molecular core as  three \emph{hard
  spheres} of radii $r_r$ located at the vertices of a
flattened isosceles triangle---large spheres in the
sketch of Fig.~\ref{fig:interaction_site}. The long base of the triangle is twice of its 
height $l$, hence all three interaction sites have equal distance $l$ from the center of the base.   
The carbonated chains of the molecules are modeled  by a small
number of sites with Lennard-Jones interactions.
Four neighboring $\text{CH}_2$ units of the alkoxychain are
coarse-grained  to  one Lennard-Jones site---small  spheres
 in Fig.~\ref{fig:interaction_site}. The interaction sites are
arranged in  straight, rigid arms with their centers separated by a distance
$\sigma$.

The radii of the hard spheres and the range of Lennard-Jones interaction, the distance of the former to
the molecular center, as well as the length and  orientation of the arms are the
parameters of the model. These geometrical features are obtained from the
 conformations which minimize molecular mechanics simulations (MM) starting
from an initial configuration derived from a
 detailed analysis of the experimental  STM images, Fig.~\ref{fig:experiment_monolayers}.
The relevant length scales of the Fr{\'e}chet dendrons
range approximately from $15\,\angstrom$ of the skeleton to $50\,\angstrom$
for the extended  molecule, with $l = 6.1\,\angstrom$,
$r_r=2.6\,\angstrom$ and $\sigma = 6\,\angstrom$.

We encode the strong intermolecular steric interaction
between the aromatic rings by hard-sphere repulsions between the three central spheres
of different molecules, preventing the cores of the  molecules from
overlapping.
The weak, short-ranged van-der-Waals attraction of the
lateral chains is described by a  Lennard--Jones potential: $V(r) =
4\,\epsilon\left[(\sigma/r)^{12}-(\sigma/r)^6\right]$ among beads in
chains of different molecules. Thus $\epsilon$ corresponds to the minimum energy which occurs at $r = 2^{1/6} \sigma$. 

The atomically flat graphite surface which
 constitutes the  template for the pattern
formation, offers
six energetically equivalent orientations for  the molecule. The molecule-substrate attraction is mainly mediated by
$\pi$-interactions between the phenyl-rings and the chains with the graphite
surface, and the total adsorption energy is about ten times larger than the total intermolecular interaction.
In addition,  as the size of the entire molecule,  around
$45\,\angstrom$, is up to ten times larger than
the lattice constant (a few $\angstrom$), only the symmetry of
the underlying substrate plays a role in the monolayer
assembly. In our modeling the molecules are fixed at the sites of a
coarse-grained, fully occupied triangular lattice. The lattice
constant  $a$ is comparable to the  size of the building blocks and the lattice
exhibits the same symmetry as the original graphite
honeycomb structure.
The Fr\'echet dendrons may   rotate by discrete angles as  rigid bodies around their
centers and adopt one of the six preferred orientations
 of the underlying graphite.

The interaction-site model accomplishes a
significant reduction of degrees of freedom, setting the flexibility
of the molecule aside.  While the
actual physical system contains hundreds of
atoms per molecule able to  displace and rotate independently, the
coarse-grained interaction-site model consists of a rigid object
with no other degrees of freedom than the rotation around its
center.

\begin{figure}[h]
\includegraphics[width=\columnwidth]{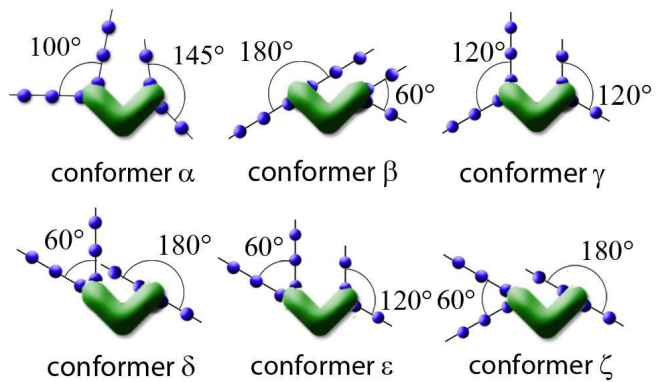}
\caption{Molecular conformations showing different orientations of the
  lateral straight arms. The angles of the arms with respect to the
  positive $x$-axis are given clockwise
  from the left to the right for the five additional conformations: 
$\beta =(-5\pi/6, 5\pi/6, 5\pi/6,
  -\pi/6)$,
$\gamma =(-5\pi/6, \pi/2, \pi/2, -\pi/6)$, $\delta=(-5\pi/6, \pi/6, \pi/2, \pi/6)$
  $\epsilon = (-5\pi/6, \pi/2, \pi/2, \pi/6)$, $\zeta = (-5\pi/6,
  \pi/6, \pi/6, -\pi/6)$.}
\label{fig:conformations}
\end{figure}

In addition to the conformation closest to the experimental findings, Fig.~\ref{fig:interaction_site}, we explore the Fr{\'e}chet dendron  
in five 
 alternative  
conformations  by varying  the
orientations of the arms, as displayed in Fig.~\ref{fig:conformations}, where the angles follow the symmetry directions of the substrate.

\subsection{Predictions of the model}
To find the regular patterns emerging in the self-organized
monolayers
we have run Monte Carlo (MC)  simulations~\cite{frenkel2002understanding,landau2005guide,binder2002monte} considering some hundred to a few thousand
molecules which corresponds also to the experimental situation: a  single
dendron covers a surface of approximately $4\,\nano\meter^2$
and the samples imaged with  STM occupy some hundred $\nano\meter^2$.
The lattice constant which  essentially fixes the packing fraction of the monolayer has been varied  from
$a=2.8\,\sigma$ to $a=4.2\,\sigma$ covering the experimental regime. 
To ensure that the stable patterns are assumed in the simulations, simulated annealing~\cite{kirkpatrick1984optimization,kirkpatrick1983optimization,cerny1985thermodynamical} has been employed. Random  configurations,
where molecules 
assume  a random direction
among the six possible orientations, have been chosen as initial configurations.
 Starting at a given
temperature  the system evolves
via Monte Carlo moves by performing $\pi/3$ rotations.
 We have  run 500 MC sweeps~\footnote{The orientation of every molecule is
   updated on average once at every MC sweep.} to
  equilibrate the sample and measure the energy and order parameters
    during another 1000 Monte Carlo sweeps before
  lowering or raising the temperature. Then
the temperature is lowered and  the process is repeated. The system is
cooled down  to temperatures around $170\;\kelvin$, low enough
for our purpose, given that the experiments were performed at room
temperature.

\begin{figure}[h]
\includegraphics[width=\columnwidth]{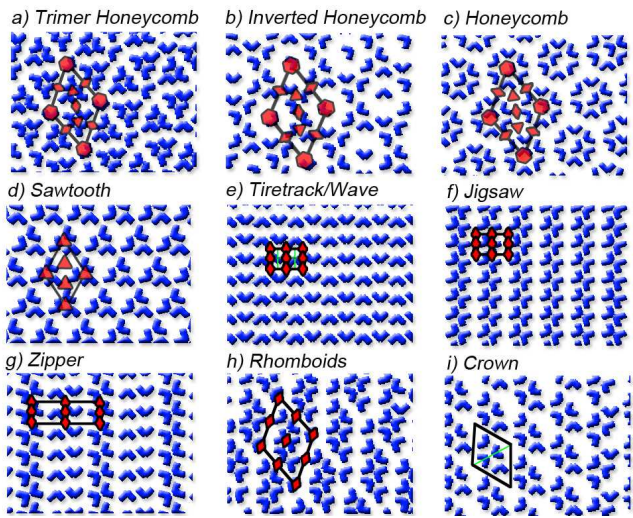}
\caption{Ordered motifs found by simulated annealing for the Fr\'echet dendrons with different conformations within
  the interaction-site model. Molecules are represented as wedges
matching their three aromatic rings.
 The carbonated arms are omitted for clarity.}
\label{fig:stable_patterns}
\end{figure}

An abundant variety of ordered patterns arises by cooling samples of
the interaction-site model in all conformations, see Fig.~\ref{fig:stable_patterns}. These patterns display different degrees of symmetries~\cite{Schattschneider:1978}  which are subgroups of the
$p6m$ symmetry (six-fold rotation, mirror symmetry) of the underlying lattice mimicking the graphite substrate. 
The honeycomb, trimer and inverted honeycomb structure, see Fig.~\ref{fig:stable_patterns}, display a $p6$ symmetry, i.e. the sixfold rotational symmetry is still present, yet the mirror symmetry of the substrate is broken. The sawtooth pattern exhibits only a $p3$ with threefold rotations of the unit cell. In the tiretrack/wave
 pattern rows of aligned
units emerge with opposite directions resulting in  a $p2gg$ symmetry, i.e. a twofold rotational symmetry and  glide-mirror axes. The class of $p2$ symmetric patterns consists
of the jigsaw, zipper and rhomboid ordering. Here the jigsaw unit cell is comprised only of two units arranged in a head-to-head configuration, whereas the more exotic zipper
contains four molecules in a unit cell, the rhomboid pattern even six building blocks. The lowest symmetry $cm$ consisting of a mirror axis and a glide mirror axis, yet no rotational symmetry, is realized in the crown pattern, where four molecules order in a head-in fashion. 

Table~\ref{tb:patterns} lists  the patterns found in the cooling
process for the different conformations. Some patterns turn out to be only metastable and the ground state may change 
upon varying the packing fraction. The configuration $\alpha$ which resembles most closely the experimental configuration displays a significantly larger variety of 
patterns.

To asses the stability of the emergent patterns, we have prepared perfectly
ordered configurations at low temperatures and heated them up slowly while
monitoring the evolution of the energy and
suitable order parameters. Thus, one can identify metastable phases which
directly transform into a disordered state. 
In particular, one can estimate the melting temperature of the broken symmetry phase and 
determine whether the transition is continuous or discontinuous. 

The order parameter is defined explicitly  for every pattern, based on
the fraction of molecules in a given sublattice following its preferred orientation.
Indeed, one can distinguish various sublattices in which the molecules
point in the same direction; for instance, alternating rows in
the tiretrack/wave pattern represent two different sublattices. Thus, for a
sublattice $A$ with $N_A$ molecules and preferred orientation
$\sigma^A$, the partial order parameter reads
\begin{equation}
 m_A = \frac {N_A^{\sigma_A}}{N_A} - \frac 1 5 \sum_{\sigma_i\neq
   \sigma_A}\frac{N_A^{\sigma_i}}{N_A}.
\end{equation}
The factor $1/5$ is introduced to ensure that $m_A=0$ in a disordered phase
where all six orientations are equiprobable.
The global order parameter we monitor is just the average of the order
parameters $m_i$
over all sublattices~$n_s$
\begin{equation}
m = \sum_{i=1}^{n_s} \frac {m_i}{n_s}.
\end{equation}
The thermal fluctuations of the order parameter encode the linear response of the system with respect to a fictitious external
aligning field. Then a susceptibility can be defined as 
\begin{equation}
\chi_m = \frac{N}{k_B T} \left(\left<m^2\right>-\left<m\right>^2\right).
\end{equation}
and its behavior as a function of temperature is indicative of the nature of the transition. Similarly, we measure in the Monte Carlo simulation
the average energy per molecule $\langle u \rangle$ and the corresponding fluctuations 
\begin{equation}c_N = \frac{N}{k_B T^2}
  \left(\left<u^2\right>-\left<u\right>^2\right),
\end{equation}
which represents the specific heat per particle.  

\begin{figure}[h]
\centering
\includegraphics[width=\columnwidth]{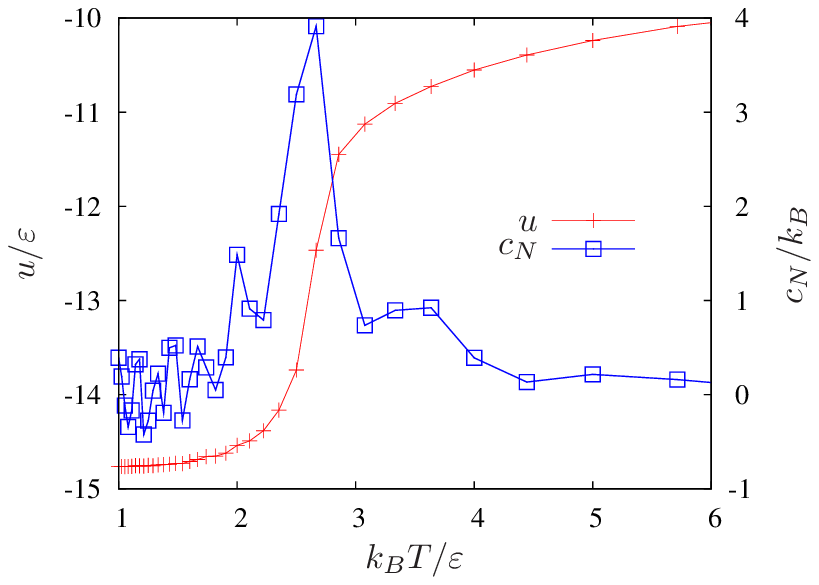} \\
\includegraphics[width=\columnwidth]{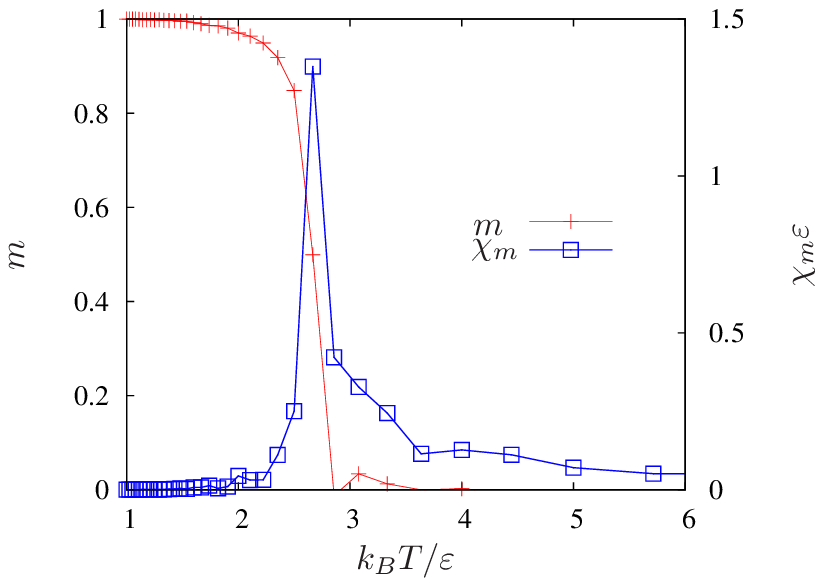}
\caption
{Melting transition for the
  sawtooth phase. Top: Energy  and heat capacity per molecule as a function of the
  dimensionless temperature. Botom: Order parameter and suscecptibility.
The simulation was performed
  for $N=576$   Fr\'echet dendrons in the $\beta$ conformation
  on a lattice with  $a/\sigma = 3.8$.}
\label{fig:exemplary_transition}
\end{figure}

The simulation results are exemplified in Fig.~\ref{fig:exemplary_transition}  for the melting transition of the sawtooth phase
as a function of the reduced temperature
$k_BT/\epsilon$.  The average energy and the order parameter drop in a small temperature interval  suggesting a first order melting transition. The corresponding
susceptibilities exhibit corresponding peaks close to the transition temperature, and are interpreted as smeared delta functions. Yet, we cannot exclude a continuous transition, and finite size scaling would be needed to clarify the order of the transition. In this work we focus on the pattern diversity and have not pursued this issue
any further.

\begin{table}[h]
\centering
\begin{tabular}{ll}
\hline\hline 
Configuration & Patterns \\
\hline
\hspace{0.5cm} $\alpha $ \hspace{1cm} & {\bf Tiretrack/wave} \\
& {\bf Sawtooth} \\
& {\bf Jigsaw} \\
& {\bf Honeycomb} \\
& Rhomboid \\
\hline
\hspace{0.5cm} $\beta$\hspace{1cm}  & {\bf Sawtooth} \\
& {\bf Tiretrack/wave} \\
\hline 
\hspace{0.5cm} $\gamma$\hspace{1cm}  & {\bf Tiretrack/wave} \\
& Honeycomb \\
& inverted Honeycomb  \\
\hline 
\hspace{0.5cm} $\delta$\hspace{1cm}  & {\bf Trimer Honeycomb} \\
\hline  
\hspace{0.5cm} $\epsilon$\hspace{1cm} & {\bf Jigsaw} \\
& {\bf Trimer Honeycomb} \\
& Crown \\
\hline 
\hspace{0.5cm} $\zeta$\hspace{1cm} & {\bf Tiretrack/wave} \\
& Zipper \\
\hline 
\end{tabular}
\caption{Emerging ordered phases by cooling down samples with $N=576$
 Fr\'echet dendrons for the  chain conformations
  $\alpha,\ldots, \zeta$.  The phases in
bold font are stable under heating, while the other phases would be unstable in an experiment at room temperature.}
\label{tb:patterns}
\end{table}

\subsection{Comparison to experiments}

The interaction-site model reproduces many of the features of the experimental model system consisting of  Fr{\'e}chet 
dendrons on graphite surfaces. First, in both systems an ample diversity of patterns emerges suggesting that the basic building blocks are
correctly transferred from the Molecular Mechanics calculations to the interaction-site model. Furthermore, the patterns found in the theoretical model closely
resemble the experimental ones both in its global and local ordering motifs, i.e. the structure displays the same wallpaper group~\cite{Schattschneider:1978}
 and a similar arrangement of the molecules in a unit cell. 

In the highly symmetric trimer honeycomb structure three molecules align in a subunit facing each other. The resulting trimer exhibits a threefold rotational symmetry without
additional mirror symmetry implying that the trimer is chiral. These trimers order in sixfold symmetric arrangement where a single molecule of unspecified orientation
resides at its  center, see Fig.~\ref{fig:stable_patterns}a). In the experimental system precisely the same structure has been observed where the center of the hexagons
appear as blurred regions in the STM measurements~\cite{Rohr:2010}. Thus it appears that these unpaired molecules are free to change their orientation rapidly, much faster than
the time resolution of the STM~\cite{Merz:2005}. 

The tiretrack/wave pattern is characterized by alternating columns of molecules pointing in the same direction. Then in the corresponding rows of this pattern 
 every second molecule possesses the same orientation. Depending on the conformation of the arms of the molecule, the pattern is formed by strong intracolumn interactions with moderately coupled columns,
or by strongly linked rows that then arrange in a parallel fashion.  Similar to the theoretical model, the experimental  system displays a tiretrack and a wave pattern where 
molecules arrange in rows or columns, respectively, see Fig.~\ref{fig:experiment_monolayers}e. Experimentally the coupling in the rows and columns is significantly different resulting into two clearly distinguishable phases,  although wave and tiretrack belong to  the same wallpaper group. For the simulations on a coarse-grained lattice this distinction can no longer be made and both phases merge into a single
tiretrack/wave pattern.

 Our simulations also reproduce the sawtooth pattern, Fig.~\ref{fig:stable_patterns}d, which has been observed in STM measurements on Fr{\'e}chet dendrons. Here three molecules form a composite which constitutes the repeat unit on a triangular lattice. In contrast the experiments report a lower wallpaper group, $p2$ rather than $p3$, 
since here two trimers arrange in a opposite orientation to form a unit cell, compare Fig.~\ref{fig:experiment_monolayers}.  

For the case of the jigsaw pattern, Fig.~\ref{fig:stable_patterns}f, we find almost perfect agreement with the STM images, both with respect to the local ordering as well
as with respect to the wallpaper group. 

The remaining patterns we have generated in the interaction-site model have not been found for Fr{\'e}chet dendrons. The first group of patterns (Fig.~\ref{fig:stable_patterns}a-c) are all highly symmetric and very similar in their respective local motifs and therefore these structures may be very sensitive to 
the conformation and the details of the molecule. Thus it appears promising to modify the chemical structure of the Fr{\'e}chet dendron only slightly to realize also
the inverted honeycomb and the honeycomb structure.  The second group of simulated patterns without corresponding experimental result
consists of large complex unit  cell with low symmetry (Fig.~\ref{fig:stable_patterns}g-i). Therefore even if some of these patterns constitute the ground state of the
system it is likely that they are not realized experimentally due to kinetic barriers.

\begin{figure}[h]
\includegraphics[width=\columnwidth]{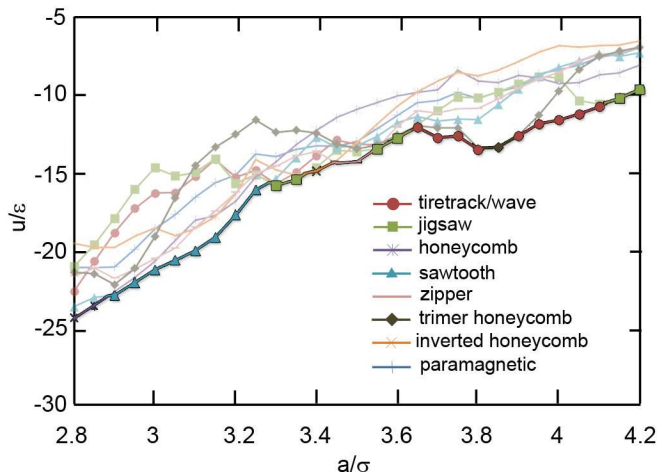}
\caption{Zero temperature energies for conformation $\gamma$ as a function of the
   lattice constant. The ground state corresponding to the lowest energy depends sensitively on the packing fraction demonstrating that the competing ordering motifs display similar energies.  
}
\label{fig:ground_state_energies}
\end{figure}

The energies of the various patterns are rather similar in agreement with the experimental observation of coexisting metastable phases. After annealing in the experiments or sufficiently many Monte Carlo steps in the simulation the patterns are expected to transform to the stable pattern. For temperatures  relevant in the experiment these patterns are almost always identified with the ground state of the system. Which of the suggested patterns actually represents the ground state depends critically on the density. We exemplify the subtle interplay of the effects of packing and ordering for conformation $\gamma$ in Fig.~\ref{fig:ground_state_energies} which displays one of the most complex behaviors. For high packing fractions (small lattice constant) first the honeycomb pattern and then the sawtooth pattern corresponds to the ground state. 
For moderate densities the tiretrack/wave exhibits the lowest energy and experimentally it also is found as the most stable pattern.

We have shown that  interaction-site models 
 are a powerful approach to model pattern formation of complex supramolecules on  a substrate. In particular, the pattern diversity is correctly transferred 
from the molecular mechanics minimizations and also details of the patterns such as symmetries and local ordering motifs are surprisingly well reproduced. Let us discuss
limitations and possible improvements of the current model. First, the interaction-site model so far accounts  only for the dominant van der Waals interactions, yet on 
closer inspection electrostatic interactions could be also important for the selection of a particular motif. From the molecular mechanics calculations one infers that
these forces are responsible for up to 30\% of the intermolecular interaction. It would be desirable to model also these electrostatic interactions and test if more accurate results can be achieved. Similarly, in the case of highly directional forces, such as hydrogen bonds, one should explicitly incorporate these specific interactions in the interaction site model.
Second,  the   model does not account for the effects of the  solvent.  Nevertheless 
some solvents interact via hydrogen bonds with the  Fr\'echet dendrons and may favor conformations of the Fr{\'e}chet dendrons where oxygen atoms are exposed to the solvent. Since the conformations are crucial for the corresponding patterns one should also account for the solvent effect in the interaction-site model. For example, one could attribute an additional conformational energy for each molecule and then compare the total energy, consisting of both conformational as well configurational energy, 
and thus determine the stable pattern and stable conformation in a specific solvent. Furthermore the lattices considered so far are fully occupied,  which 
excludes complex motifs with host structures. In the interaction-site model one could account for this possibility by either a fixed partial coverage density (canonical ensemble) or a fixed chemical potential (grand canonical ensemble) and include hopping processes in addition to the rotational Monte Carlo moves. Last, we have reduced the molecule-substrate interaction to the lattice symmetry and spacing. The major justification for this simplification is that the supramolecules are significantly larger
than a unit cell of the substrate and correspondingly experience only an averaged adsorption energy with preferential directions. Yet, for smaller molecules, larger
substrate lattice constants, or specific interactions one should include explicit space-resolved molecule-substrate interactions.

\section{Conclusions and outlook}

In this work, we have developed an interaction-site model which
correctly predicts the ordered
motifs of assembled monolayers. By reducing the degrees of freedom and considering the
building-blocks as rigid bodies with a reduced set of interacting
points, we have demonstrated that the self-organization relies on very
general features of the system considered:
the coupling with the substrate, the
geometry of the building blocks, and their weak interactions.
These are universal
principles in self-organizing  systems which do not
depend on the specific nature of the building blocks and the underlying
substrate---our method also works for  substrates exhibiting
different geometries. Therefore, the predictive power provided by our model
 may  guide the synthesis of suitable building blocks to engineer
arbitrary patterns for specific goals. The versatility offered to construct
the   building blocks  makes our model especially suitable to explore a  wide range of geometries.

In addition, we have shown that by combining simulated annealing with
a subsequent slow heating of the system, one can not only predict the
emergent patterns, but also their stability upon heating and the
nature of the transition into a disordered phase. 
 We have found that a broad
variety of long-range ordered phases are stable for various
conformations of the building blocks and density regimes which may
indeed coexist, as it has been observed in the experiments. The melting
temperatures of the ordered motifs range from approximately
$500\,\kelvin$ to  $1500\,\kelvin$, much higher than the room
temperature where the experiments were performed. 

However, the model still lacks an active determination of the
intramolecular conformation. Here we have
considered the building blocks as rigid bodies with the conformations
observed in the molecular mechanics minimizations.
A natural generalization  would be to make the building blocks
flexible and explore the interplay between intra and inter-molecular
ordering. This extension implies including at least four new degrees of
freedom per molecule, the orientations of the arms, which  makes the Monte Carlo
simulations computationally very expensive. An interesting alternative
to speed up simulations would be to rely on  genetic algorithms~\cite{stephanie1993genetic,holland1992genetic}.
They are suitable methods
 to compute the ground states of two-dimensional systems by minimizing the energy of single unit
cells~\cite{Fornleitner:2008,Gottwald:2005}. The conformations and
motifs resulting from this minimization can serve as input
configurations for the Monte
Carlo simulations to assess their stability.  In this way, the interaction-site
model is less dependent on the details of the  external input.
In addition, when employing  genetic algorithms one can   relax the constraint
of the substrate. Instead considering the building blocks to be
attached to the sites of a lattice, one can mimic the role of the substrate
through arbitrary potentials. This opens a way to investigate systems whose substrates
 display  more complex symmetries and interactions with the building-blocks.

\section{Acknowledgements}
We thank Erwin Frey stimulating discussions. 
MBG gratefully acknowledges the financial support of La Caixa and
DAAD. C.R. has been supported by the ``Studienstiftung des deutschen Volkes`` and the ``Elite Network Bavaria``. 
This project is supported by the  German Excellence Initiative via the program ``Nanosystems Initiative Munich (NIM).''

%

\end{document}